\documentstyle[twoside,fleqn,espcrc2,epsf]{article}

\newcommand{\AmS}{{\protect\the\textfont2
  A\kern-.1667em\lower.5ex\hbox{M}\kern-.125emS}}

\title{ Particle Physics Implications of Neutrinoless Double Beta
Decay\footnote{Invited talk presented at the Neutrino98 conference,
Takayama, Japan; June, 1998 ( to appear in the proceedings )} }

\author{R. N. Mohapatra\address{Department of Physics, University of 
Maryland, College Park, MD-20742, U. S. A.}%
\thanks{Work supported by the National Science Foundation Grant 
No.PHY-9802551}}

\begin{document}

\begin{abstract}
Neutrinoless double beta decay is a sensitive probe of the patterns of 
neutrino masses and mixings if the neutrinos are Majorana particles as 
well as other new physics scenarios beyond the standard model. In this 
talk, the present experimental lower
bound on the lifetime for $\beta\beta_{0\nu}$ is used to
constrain the neutrino mixings and set limits on the 
parameters of the new physics scenarios such as the left-right symmetric 
models, R-parity violating SUSY models etc which lead to neutrinoless double 
beta decay. We then discuss proposed high precision searches for 
$\beta\beta_{0\nu}$ decay that can provide extremely valuable insight
not only into the nature of neutrino mixings and masses but also
put constraints on (or even rule out) new physics scenarios.  
 \end{abstract}

\maketitle

\section{Introduction}

In the standard electroweak model of Glashow, Weinberg and Salam, the
absence of the right-handed neutrinos and the existence of an exact accidental
global $B-L$ symmetry guarantees that the neutrinos are  massless to
all orders in perturbation theory. Any experimental evidence for
a non-zero neutrino mass therefore constitutes evidence for
new physics beyond the standard model and will be a major step
towards a deeper understanding of new forces in nature\cite{book}.
Among the many experiments that are under way at this moment 
searching directly or indirectly (e.g. via neutrino 
oscillations) for neutrino masses, one of the most important ones
is the search for neutrinoless double beta decay. This process is 
allowed only if 
the neutrino happens to be its own antiparticle ( Majorana neutrino)
as is implied by many extensions of the standard model. In fact there is 
a well-known theorem\cite{boris} that states that any evidence for 
neutrinoless 
double beta decay is an evidence for nonzero Majorana mass for the 
neutrinos. It is of course a much more versatile probe of new physics
as we will discuss in this article. The point is that 
 since $\beta\beta_{0\nu}$ decay changes lepton number 
($L_e$) by two units  any theory that contains interactions that violate
electron lepton number $L_e$ can
in principle lead this process. This therefore reflects the tremendous 
versatility of 
$\beta\beta_{0\nu}$ decay as a probe of all kinds of new physics beyond the
standard model. Indeed we will see that already very stringent constraints
on new physics scenarios such as the left-right symmetric models
with the see-saw mechanism\cite{MS} and
supersymmetric models with R-parity violation\cite{Rp},
scales of possible compositeness of leptons etc
are implied by the existing experimental limits\cite{klap} on this process. 
For a more detailed discussion of the theoretical situation than is 
possible here, see \cite{moh}. For an update of the experimental situation,
both ongoing and in planning stage, see \cite{klap2}.

This talk is organized as follows:
 In section 2, I discuss the basic mechanisms for neutrinoless
double beta decay ; in section 3, the implications of the present limits on 
the lifetime for neutrinoless double beta decay for neutrino mixings 
are discussed; in part section 4, I go on to discuss the kind of new physics 
scenarios that can be probed by $\beta\beta_{0\nu}$ decay and the  
constraints on the parameters of the new physics scenarios
implied by present data.

\section{ Mechanisms for $\beta\beta_{0\nu}$ decay}

As is wellknown, if the neutrino is its own antiparticle, the conventional
four-Fermi interaction can lead to neutrinoless double beta decay via the
diagram in Fig. 1. In physics
scenarios beyond the standard models, if there are heavy Majorana 
fermions interacting with the electrons, diagrams similar to Fig. 1 with 
neutrino line replaced by the Majorana fermions can also lead to
$\beta\beta_{0\nu}$ decay. Examples of such particles abound in literature:
right-handed neutrino, photino, gluino to mention a few popular ones.

\begin{figure}[htb]
\begin{center}
\epsfxsize=7.5cm
\epsfysize=7.5cm
\mbox{\hskip -1.0in}\epsfbox{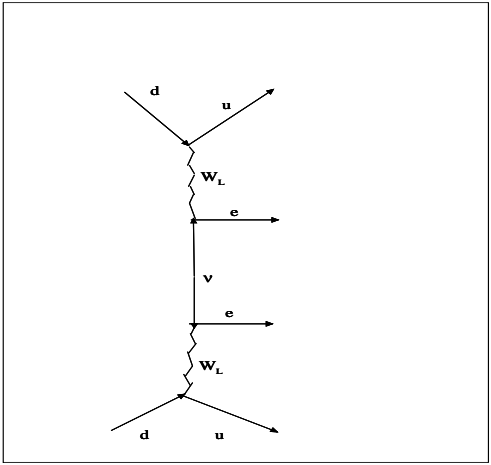}
\caption{ Feynman diagram involving neutrino majorana mass
that contributes to $\beta\beta_{0\nu}$ decay.
\label{Fig.1}} 
\end{center}    
\end{figure}

One could therefore give an arbitrary
classification of the mechanisms for $\beta\beta_{0\nu}$ decay into two
kinds: (A) one class that involves the exchange of light neutrinos; and 
(B) the second class that involves heavy fermions or bosons. 
Furthermore, there 
are two distinct mechanisms for light neutrino exchange contributions:
(a) helicity flip light neutrino mass mechanism and
(b) helicity nonflip vector-vector or vector-scalar mechanism.
In case (a), one can write the amplitude
$A_{\beta\beta}$ for neutrinoless double beta decay 
to be:
\begin{equation}
A^{(m)}_{\beta\beta}\simeq {{G^2_F}\over{2}}\langle{{m_{\nu}}\over{k^2}}\rangle
_{Nucl.}
\end{equation}
whereas in case (b), it looks like:
\begin{equation}
A_{\beta\beta}\simeq {{G^2_F}\over{2}} \langle 
{{\eta}\over{\gamma\cdot k}}\rangle_{Nucl.}
\end{equation}
\vspace{6mm}
To extract neutrino mass implications for neutrinoless double beta decay, we
need to note the explicit form of $<m_{\nu}>$:
\begin{equation}
<m_{\nu}>~ = \Sigma_{i} U^2_{ei} m_i
\end{equation}
where $U_{ei}$ are the mixing matrix elements for the electron neutrino
with the other neutrinos. Therefore a constraint on the $<m_{\nu}>$ can be
converted into constraints on the neutrino mixings involving the first 
generation. Incidentally, one can also write $<m_{\nu}>= m_{ee}$ where
$m_{ee}$ is the $ee$ entry of the neutrino mass matrix in the weak basis.
Thus any theory which has zero entry in the $ee$ location leads to
vanishing neutrinoless double beta decay even if the neutrino is a 
Majorana particle.

It is important to remark that these kind of light neutrino exchange
diagrams always lead to a long range neutrino potential inside the
nucleons and therefore, crudely speaking the two nucleons "far" from
each other can contribute in the double beta decay. This has 
important implications
for the evaluation of the nuclear matrix element\cite{stoica}, an important 
subject we do not discuss here. We will instead use an effective momentum 
to parameterize the effect of the nuclear matrix 
element calculations (we will roughly choose $p_{eff}\approx 50$ MeV).
The width for double 
beta decay amplitude is given by
\begin{equation}
\Gamma_{\beta\beta}\simeq {{Q^5|A|^2}\over{60\pi^3}}
\end{equation}
Here, $Q$ is the available energy for the two electrons.
 Using the present most stringent limit on 
$\tau_{\beta\beta}\geq 1.1\times 10^{25}$ years obtained for $^{76}Ge$
by the Heidelberg-Moscow group, one can obtain the upper limit on
the width to be
$\Gamma_{\beta\beta}\leq 3.477\times 10^{-57}$ GeV; using Eq. (4), $A$ 
for the light neutrino contribution,
 $Q\simeq 2$ MeV and $p_F\simeq 50$ MeV, 
one gets a rough upper limit of .7 eV for the neutrino mass.  
A more careful estimate leads to
\begin{equation}
<m_{\nu}>\leq .46~ eV \nonumber\\
\eta \leq 10^{-8}
\end{equation}

\noindent (B): The second class of mechanisms consists of exchange of heavy
particles which often arise in physics
scenarios beyond the standard model. In the low energy limit, the
effective Hamiltonian that leads to $\beta\beta_{0\nu}$ decay in these
cases requires point interaction between nucleons; as a result, in general
the nuclear matrix elements are expected to be smaller
due to hard core repulsive nuclear potential;
nevertheless, a lot of extremely useful information have been extracted
about new physics where these mechanisms operate. Symbolically, such 
contributions can arise from effective Hamiltonians of the following
type( we have suppressed all gamma matrices as well as color indices):
\begin{equation}
H^{(1)}~=G_{eff}~\overline{u}\Gamma d \overline{e}\Gamma F~+~h.c.
\end{equation}
or
\begin{equation}
H^{(2)}~=\lambda_{\Delta}\left({\frac{1}{M^3}}~\overline{u}\Gamma d
\overline{u}\Gamma d~+~e^-e^-\right)\Delta^{++} \nonumber\\
+~h.c.
\end{equation}

Here $F$  represents a neutral majorana fermion such as the right-handed
neutrino ($N$)\cite{rosen} or gluino $\tilde{G}$ or photino $\tilde{\gamma}$ and
  $\Delta^{++}$ represents a doubly charged scalar or vector particle. 
In the above equations, the coupling $G_{eff}$ has dimension of $M^{-2}$
and $\lambda_{\Delta}$ is dimensionless.
The possibility of the doubly charged scalar contribution to 
$\beta\beta_{0\nu}$ was first noted in \cite{MV} and have been discussed
subsequently in \cite{SV}.
The contributions to neutrinoless double beta decay due to the
above interactions lead 
to $\beta\beta_{0\nu}$ amplitudes of the form:
\begin{equation}
A^{(F)}_{\beta\beta}\simeq ~G^2_{eff}{{1}\over{M_F}}(p^{eff})^3
\end{equation}
and 
\begin{equation}
A^{\Delta}_{\beta\beta}\simeq ~\left({{\lambda^2_{\Delta}}\over{M^3 M^2_{\Delta}}}
\right)(p^{eff})^3
\end{equation}

Here again we have crudely replaced all nuclear effects by the effctive
momentum parameter $p^{eff}$. If we choose $p^{eff}\simeq 50$ MeV, then
the present lower limit on the lifetime for $^{76}Ge$ decay leads to a
crude upper limit on the effective couplings as follows:
\begin{equation}
G_{eff}\leq 10^{-7}\left({{M_F}\over{100~GeV}}\right)^{{1}\over{2}} 
\end{equation}
and
\begin{equation}
\lambda_{\Delta}\leq 10^{-3}\left({{M}\over{100~GeV}}\right)^{{5}\over{2}}
\end{equation}

In the second equation above, we have set $M=M_{\Delta}$. Note that these
limits are rather stringent 
and therefore have the potential to
provide useful constraints on the new physics scenarios that lead to such
particles.

\section{Implications for neutrino masses and mixings}

This conference watched the history of neutrino physics take a remarkable 
new turn. Convincing evidence was presented by the Super-Kamiokande 
collaboration for the existence of neutrino oscillation of the 
atmospheric muon neutrinos to either $\nu_{\tau}$ or a sterile neutrino.
Using data both in the sub-GeV and
multi-GeV energy range for the electron and the muon neutrinos as well as 
the zenith angle 
dependence of the muon data, the present fits at 90\% confidence level 
seem to imply the following values for the oscillation parameters
$\Delta m^2$ and $sin^22\theta$:
$4\times 10^{-4}\leq \Delta m^2\leq 5\times 10^{-3}$ eV$^2$ with
$sin^22\theta$ between .8 to 1\cite{Kajita}.
The possibility of $\nu_{\mu}-\nu_{e}$ oscillation as an explanation of 
the atmospheric anomaly seems to run into conflict with the recent CHOOZ
\cite{CHOOZ} experiments.
Neutrino oscillation also seems to be the only way to understand the deficit
of the solar neutrinos\cite{bahcall}. The detailed oscillation mechanism 
in this case is however is unclear. The three possibilities are: 
   { a)}{ Small-angle MSW\cite{MSW}, }$\Delta m^2_{ei}\simeq 6\times10^{-6}{ 
eV}^2,
         sin^22\theta_{ei}\simeq 7\times10^{-3}$;
  { b)}{ Large-angle MSW, }$\Delta m^2_{ei}\simeq 9\times10^{-6}{eV}^2,
         sin^22\theta_{ei}\simeq 0.6$;
  { c)}{ Vacuum oscillation, }$\Delta m^2_{ei}\simeq 10^{-10}{eV}^2,
         sin^22\theta_{ei}\simeq 0.9$.
The data on neutrino energy distribution presented at this conference
indicates a preference towards vacuum oscillation rather than MSW mechanism.
Turning to the laboratory experiments, the LSND\cite{LSND} collaboration has
presented evidence in favor of a possible oscillation
of $\overline{\nu}_\mu\rightarrow \overline{\nu}_e$ as well as 
$\nu_{\mu}-\nu_e$. The preferred $\Delta m^2$ range seems to be
$.24 \leq \Delta m^2_{e-\mu}\leq 10$ eV$^2$ with a mixing angle in the
fea percent range.  As already mentioned, $\beta\beta_{0\nu}$ gives 
only an upper bound of $<m_{\nu_e}>< .46\,$eV.

Another effect of neutrino mass is in the arena of cosmology, where it 
not only effects whether the universe keeps expanding for ever or it 
eventually collapses onto itself, but it also determines the detailed 
manner in which structure formed in the early universe. This subject
is in a constant state of flux due to new cosmological data coming in
at a very rapid rate. But the idea that the present structure data
may need a neutrino mass contribution to the dark matter is very much alive
(see for instance Ref.\cite{silk} which seems to suggest that a total
neutrino mass of 4-5 eV which contributes about 20\% of the dark matter 
along with 70\% cold dark matter and 10\% baryon gives the best fit to
the galaxy power spectrum data. This taken seriously would mean that
$\Sigma_i m_{\nu_i}= 4 - 5 $ eV).

 With the above input information, if we stay within the minimal
three neutrino picture, then the solar neutrino puzzle can be 
resolved by $\nu_e\to\nu_\mu$ oscillations and the atmospheric
neutrino deficit by $\nu_\mu\to\nu_\tau$ oscillations and the LSND 
results cannot be accomodated. 
Note that these observables are controlled only by the mass square 
difference; on the other hand, the required hot dark
matter implies that at least one or more of the neutrinos must have 
mass in the few eV range. It was pointed out\cite{calmoh} in 1993 that,
 in the minimal picture, this leads to a scenario, where
all three neutrinos are nearly degenerate, with $m_{\nu_e}\approx 1.6$ eV.
It is then clear that, in this case, in general there will be an observable 
amplitude for neutrinoless double   
beta decay mediated by the neutrino mass. In fact, if the 
limit on $\langle m_\nu\rangle$ is taken to be less than $.47$ eV  
as is implied for a certain choice of the nuclear matrix element,
then the mixing must satisfy the constraint:
\begin{eqnarray}
\Sigma_i U^2_{ei}\simeq 0
\end{eqnarray}
Since each of the elements in the above sum is complex, the $U^2_{ei}$ 
form the three sides of a triangle\cite{vissani}. Then using the 
unitarity relation for the $U$ matrix, it is clear that one must
have $|U_{ei}|\leq 1/2$. On the other hand, the CHOOZ data for a general 
three neutrino oscillation picture implies that 
$4|U_{e3}|^2(1-|U_{e3}|^2\leq .2$. These two constraints then imply that
$|U_{e3}|\leq .2$. This is indeed an interesting constraint and rules out
(provided of course $\Delta m^2_{e3} \geq 10^{-3}$ eV$^2$) a maximal 
mixing scenario for degenerate neutrinos that was proposed to reconcile 
sub-eV double beta decay neutrino mass limit i.e.\cite{nuss}
\begin{eqnarray}
U= \frac{1}{\sqrt{3}}\left(\begin{array}{ccc}
1 & \omega & \omega^2\\
1 & \omega^2 & \omega\\
1 & 1 & 1 \end{array}\right)
\end{eqnarray}

There is however another mixing pattern for the 
degenerate neutrino scenario which is consistent with both the CHOOZ 
experiment
and the neutrinoless double beta decay bounds:
\begin{eqnarray}
U=\left(\begin{array}{ccc}
1/\sqrt{2} & i/\sqrt{2} & 0 \\
1/\sqrt{6} & -i/\sqrt{6} & -2/\sqrt{6} \\
1/\sqrt{3} & -i/\sqrt{3} & 1/\sqrt{3}
\end{array}\right)
\end{eqnarray}
Other more general constraints for this case have been studied in several 
recent papers\cite{mina}.

If we do not include the hot dark matter constraint, then there is no 
need to require that the neutrinos are degenerate in mass and one can 
live perfectly happily with a hierarchical pattern of neutrino masses as 
dictated by the simple type I seesaw formula. In that case, one can 
combine the atmospheric oscillation fits and the CHOOZ data to set an 
upper limit on $<m_{\nu}>$ equal to $\sqrt{\Delta m^2_{ATMOS}} 
sin^2\theta_{e\tau}\simeq .02$ eV\cite{bile}. Thus evidence for $<m_{\nu}>$ 
above 
this value would be an indication that either the neutrino mass pattern 
is not hierarchical or that the atmospheric neutrino puzzle involves 
transition between $\nu_{\mu}$ and a sterile neutrino. Both of these are
extremely valuable conclusions. The GENIUS proposal of the Heidelberg 
group\cite{klap1} is expected to push the double beta decay limit to this 
level and could therefore test this conclusion.

\section{ Implications for physics beyond the standard model:}

Let us now discuss the constraints implied by neutrinoless double beta 
decay searches on the new physics scenarios 
beyond the standard model. Let us first consider
the the neutrino mass mechanism. Any theory which gives the electron
neutrino a significant ( $\simeq$ eV ) Majorana mass or any other species
( e.g. $\nu_{\mu}$ or $\nu_{\tau}$ ) a large enough mass and 
mixing angle with the $\nu_e$ so that $U^2_{ei}m_{\nu_i}$ is of order
of an electron volt will make itself open to testability by the 
$\beta\beta_{0\nu}$ decay experiment. There are many theories with
such expectations for neutrinos. Below I described two examples: (i) the
singlet majoron model and (ii) the left-right symmetric model.
Both these models are intimately connected with
ways to understand the small neutrino mass in gauge theories.

\subsection{ The singlet majoron model:}

This model\cite{cmp} is the simplest extension of the standard model
that provides a naturally small mass for the neutrinos by employing the
the see-saw mechanism\cite{seesaw}. It extends the
standard model by the addition of three right-handed neutrinos and the
addition of a single complex Higgs field $\Delta$
 which is an $SU(2)_L\times U(1)_Y$
singlet but with a lepton number +2. There is now a Dirac mass for the 
neutrinos and a Majorana mass for the right handed neutrinos proportional
to the vacuum expectation value (vev) $\langle\Delta\rangle\equiv v_R$.
This leads to a mass matrix for the neutrinos with the usual see-saw form:
\begin{equation}
M=\left(\begin{array}{cc}
{0}&{m_D}\\{m^T_D}&{fv_R}\end{array}\right)
\end{equation}

This leads to both the light and heavy (right-handed) neutrinos being
Majorana particles with the mutual mass relation being given by the
see-saw formula:

\begin{equation}
m_{\nu_i}\simeq{{m_{iD}(M^{-1}_{iR})m^T_{iD}}}
\end{equation}
where we have ignored all mixings and $M_{iR}\simeq f_{ii}v_R$
denote the masses of the heavy right-handed neutrinos . It is clear that
the electron neutrino mass can be in the electron-volt range if the 
values of $m_{1D}$ are chosen to be of similar order of magnitude to
the electron mass. In fact, for $m_{1D}= m_e$, and $m_{1R}=250$ GeV,
one gets $m_{\nu_e}=1$ eV which is the range of masses being probed
by the ongoing and proposed $\beta\beta_{0\nu}$ experiments. 

More importantly, this class of models leads to the new neutrinoless double 
beta decay process with majoron emission\cite{nuss1} which has a very 
different electron energy distribution than either $0\nu$ or $2\nu$ 
double beta decays. The relevant Feynman diagram is same that in Fig. 1 with
a majoron line emanating from the light neutrino in the middle. The 
majoron coupling $g_{\nu\nu\chi}$ then replaces the neutrino mass in the 
$\beta\beta_{0\nu}$ amplitude. 
This observation has led to a considerable amount of experimental effort 
into searching for the majoron emitting double beta decay and limits at the
level of $g_{\nu\nu\chi}\leq 10^{-5}$ are presently available.

A relevant question is whether majoron couplings at the level 
measurable are expected in reasonable extensions of the standard model. 
There have been extensive studies of this question and is beyond the 
scope of this review. But it is of interest to note that in the simplest
singlet majoron model, one expects $g_{\nu\nu\chi}\simeq \Sigma_a m^2_{ea}
M^{-2}_a g_{aa\chi}$. In the absence of any mixings, this is proportional to
$m_{\nu_1}/M_{N_1}$ which is expected to be of order $10^{-11}$ for an eV 
$\nu_e$ and 100 GeV for the $B-L$ breaking scale. However, if the 
$\nu_{\tau}$ mass is in the MeV range as is allowed by LEP analysis, this 
coupling could easilly be in the $10^{-5}$ to $10^{-6}$ range which is 
clearly in the range accessible to experiments.

\subsection{ Left-right symmetric models:}

Let us now consider 
the minimal left-right symmetric model with 
a see-saw mechanism for neutrino masses as
described
in \cite{MS}. Below, we 
 provide a brief description of the
structure of the model.
The three generations of quark and
 lepton fields are denoted by $Q^T_a\equiv ( u_a,d_a ) $
and $\Psi^T_a \equiv (\nu_a,~  e_a  )$ respectively,
where $a~ =~ 1,~ 2,~ 3$ is the generation index.
 Under the
gauge group $SU(2)_L \times SU(2)_R \times U(1)_{B-L}$, they are 
assumed to transform as
$\Psi_{a~L} \equiv (1/2, ~ 0 , ~ -1 )$
and $\Psi_{a~R} \equiv (0, ~ 1/2, ~ -1 )$ and similarly for the quarks
denoted by $Q^T\equiv( u,~d )$.
In this model, there is a right-handed counterpart to the $W^{\pm}_L$
to be denoted by $W^{\pm}_R$. Their gauge interactions then lead to
the following expanded structure for the charged weak currents
in the model for one generation prior to symmetry breaking
 ( for our discussion , the quark mixings
and the higher generations are not very important; so we will ignore them
in what follows.)
\begin{equation}
L_{wk}={{g}\over{2\sqrt{2}}}[W^{-}_{\mu L}J^{\mu}_L
+ L\rightarrow R]
\end{equation}
where $J^{\mu}_L=
\left(\overline{d}\gamma^{\mu}(1-\gamma_5)u
+\overline{e}\gamma^{\mu}(1-\gamma_5)\nu_e\right)$

 The Higgs sector 
of the model consists
of the bi-doublet field
$\phi \equiv (1/2, ~ 1/2, ~ 0)$ and triplet Higgs fields:
${
\Delta_L ( 1, ~0, ~ +2 ) \oplus \Delta_R (0, ~ 1, ~ +2 )
~~~~~.}$

The Yukawa couplings for the lepton sector
which are invariant under gauge and parity symmetry can be written as:
\begin{eqnarray}
{\cal L}_Y
=  {\overline \Psi_L} h^{\ell} \phi \Psi_R + {\overline \Psi_L}
{\tilde h}^{\ell}{\tilde \phi} \Psi_R  + \nonumber \\
 \Psi^T_L f \tau_2 {\vec \tau} \cdot {\vec \Delta_L} C^{-1} \Psi_L
                                + L\rightarrow R  + h.c.    
\end{eqnarray}
\noindent where
$h, ~{\tilde h}$ are hermitian matrices while 
 $f $ is a symmetric matrix in the generation space.  $\Psi$ and $Q$ 
here denote the
leptonic and quark doublets respectively.  

The gauge symmetry is spontaneously broken by the vacuum expectation
values: ${< {\Delta_R^0} > = v_R ~~; }$
${< \Delta_L^0 > =  0 ~~;}$ and
${< \phi > = 
\left(\begin{array}{cc}{\kappa}&{0}\\{0}&{\kappa^\prime}\end{array}\right) 
~~.}$
 As usual, $< \phi >$ gives masses to the charged fermions and Dirac masses
to the neutrinos whereas 
$< \Delta_R^0 >$
 leads to the see-saw mechanism for the neutrinos in the standard way\cite{MS}.
For one generation the see-saw matrix is in the form $m_{\nu}\simeq 
m^2_{f}/fv_R$ and 
leads as before to a light and a heavy state as discussed in the previous
section. For our discussion here it is important to know the structure
of the light and the heavy neutrino eigenstates:
\begin{eqnarray}
\nu \equiv \nu_e~+~\xi N_e \nonumber \\
N \equiv~N_e~-~\xi\nu_e
\end{eqnarray}
where $\xi\simeq \sqrt{m_{\nu_e}/m_N}$ and is therefore a small number.
Substituting these eigenstates into the charged current Lagrangian, we 
see that the right-handed $ W_R$ interaction involves
also the light neutrino with a small strength proportional to $\xi$.
To second order in the gauge coupling $g$, the effective weak interaction
Hamiltonian involving both the light and the heavy neutrino becomes:
\begin{eqnarray}
H_{wk}={{G_F}\over{\sqrt{2}}}(\overline{u}\gamma^{\mu}(1-\gamma_5)d
[\overline{e}\gamma_{\mu}
[(1-\gamma_5)\nonumber\\
+\xi({{m^2_{W_L}}\over{m^2_{W_R}}})(1+\gamma_5)]\nu
+\xi\overline{e}(1-\gamma_5)N] \nonumber \\
+{{G_F}\over{\sqrt{2}}}\left({{m^2_{W_L}}\over{m^2_{W_R}}}\right)
\left(\overline{u}\gamma^{\mu}(1+\gamma_5)d\overline{e}\gamma_{\mu}
(1+\gamma_5)N\right)\nonumber\\
+~h.c.
\end{eqnarray}

From Eq. (20),
 we see that there are several contributions to the 
$\beta\beta_{0\nu}$. Aside from the usual neutrino mass diagram ( Fig.1),
there is a contribution due to the wrong helicity admixture 
with $\eta\simeq \xi\left({{m^2_{W_L}}\over{m^2_{W_R}}}\right)$
 and there are 
contributions arising from the exchange of heavy right-handed neutrinos.
 This last contribution is given by :
\begin{equation}
A^{(R)}_{\beta\beta}\simeq {{G^2_F}\over{2}}\left({m^4_{W_L}\over{m^4_{W_R}}}
+\xi^2\right){{1}\over{m_N}}
\end{equation}

The present limits on neutrinoless double beta decay lifetime then
imposes a correlated constraint on the parameters $m_{W_R}$ and 
$m_N$\cite{moha1}. If we combine the theoretical constraints of vacuum 
stability then, 
the present $^{76}Ge$ data provides a lower limit on the masses of
the right handed neutrino ($N_e$) and the $W_R$ of 1 TeV, which is
a rather stringent constraint. We have of course assumed that the 
leptonic mixing
angles are small so that there is no cancellation between the parameters.

Finally, the Higgs
sector of the theory generates two types of contributions to 
$\beta\beta_{0\nu}$ decay. One arises from the coupling of the doubly
charged Higgs boson to electrons ( see Fig.2). The amplitude for the
decay is same as in Eq. (6) except we have $\lambda_{\Delta}=f_{11}$ and
\begin{equation}
{{\lambda_{\Delta}}\over{M^3}} 
=2^{7/4}G^{3/2}_F\left({{m_{W_L}}\over{M_{W_R}}}\right)^3 \end{equation}

Using this expression, we find that the present $^{76}Ge$ data implies
that ( assuming $m_{W_R}\geq 1$ TeV ) 
\begin{equation}
M_{\Delta^{++}}\geq \sqrt{f_{11}}~~ 80 GeV
\end{equation}
A second type Higgs induced contribution arises 
from the mixing among the charged Higgs fields
in $\phi$ and $\Delta_L$ which arise from the
couplings in the Higgs potential, such as Tr$(\Delta_L \phi \Delta_R^{\dagger}
  \phi^{\dagger})$ after the full gauge symmetry is broken
 down to $U(1)_{em}$. Let us denote this mixing term by an angle $\theta$.
This will contribute to the four-Fermi interaction
of the form given by the $\epsilon_1^{ee}$ term with 
\begin{equation}
\epsilon_1^{ee}\simeq
{{h_u f_{11} sin 2\theta}\over {4\sqrt{2}G_F M^2_{H^{+}}}}~,
\end{equation}
where we have assumed
that $H^+$ is the lighter of the two Higgs fields.  We get
$h_u f_{11}{\rm sin}2\theta 
\leq 6\times 10^{-9}(M_{H^+}/ 100~GeV)^2$, which is quite
a stringent constraint on the parameters of the theory. To appreciate
this somewhat more, we point out that one expects $h_u\approx m_u/ m_W
\approx 5 \times 10^{-5}$ in which case, we get an upper limit for the coupling
of the Higgs triplets to leptons $f_{11}{\rm sin}2\theta
\leq 10^{-4}$ (for $m_{H^+} = 100 ~GeV$).  
Taking a reasonable choice of $\theta \sim M_{W_L}/M_{W_R}
\sim 10^{-1}$ would correspond to a limit $f_{11} \le 10^{-3}$.  
Limits on this
parameters from analysis\cite{swartz}
 of Bhabha scattering is only of order $.2$ or
so for the same value of the Higgs mass.

\begin{figure}[htb] \begin{center}
\epsfxsize=7.5cm
\epsfysize=7.5cm
\mbox{\hskip -1.0in}\epsfbox{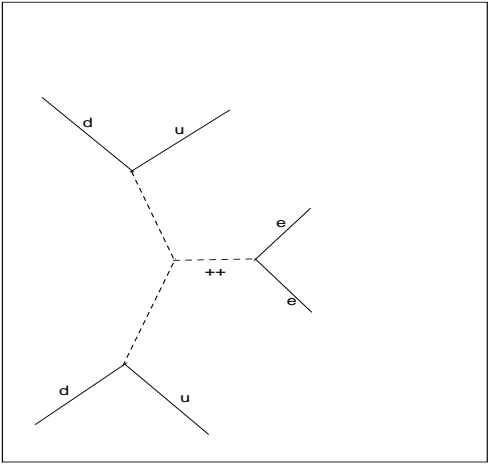}
\caption{ The Feynman diagram responsible for neutrinoless 
double beta decay due to the exchange of doubly charged Higgs bosons.
The top and bottom solid lines are quark lines and the middle 
right solid  
lines are electron lines. The dashed lines are the scalar
bosons with appropriate quantum numbers.\label{Fig.2}} 
\end{center}    
\end{figure}

An interesting recent development is that once one supersymmetrizes the 
seesaw version of the left-right model just described, allowed values for 
the right handed scale get severely restricted by the requirement that 
the ground state of the tehory conserve electric charge. There are only two
allowed domains for $M_{W_R}$: (i) if the ground state breaks R-parity,  
there is an upper limit on the $W_R$ scale of about $\leq 10$ 
TeV\cite{kuchi}. Since in this case, R-parity is spontaneously 
broken R-parity violating interactions conserve baryon number and the 
theory therefore is much improved in the sense of naturalness over the MSSM. 
What is interesting is that the GENIUS experiment can then completely 
scan the allowed range of this model. On the other hand,
if R-parity is conserved, there must be a lower limit on $M_{W_R}$ of 
about $10^{10}$ GeV\cite{chacko}. In this case also there is a 
contribution to $\beta\beta_{0\nu}$ decay coming from the light doubly 
charged Higgs boson in the same manner described above\cite{moh3}. This 
contribution scales like $V^{-2}_R$ in the amplitude. Thus as the limits 
on neutrinoless double beta decay improve, at some point they will not only
imply that the $W_R$ mass is not only bigger than $10^{10}$ GeV or so; but
they can also continue to improve this lower limit due to the 
contribution from the doubly charged Higgs boson whose mass is directly 
proportional to the square of $v_R$. 

\subsection{ MSSM with R-parity violation:}

The next class of theories we will consider is the supersymmetric
stamdard model.
As is well-known, the minimal supersymmetric standard model can have
explicit\cite{Rp} violation of the R-symmetry
(defined by $(-1)^{3B+L+2S}$), leading to lepton
number violating interactions in the low energy  Lagrangian.
The three possible types of couplings in the
superpotential are : 

\begin{eqnarray}
W^{\prime}~=~\lambda_{ijk}L_iL_jE^c_k+\lambda^{\prime}_{ijk}L_iQ_jD^c_k
\nonumber\\
+\lambda^{''}_{ijk}U^c_iD^c_jD^c_k~.
\end{eqnarray}
Here $L,Q$ stand for the lepton and quark doublet superfields, $E^c$ for
the lepton singlet superfield and 
$U^c,D^c$ for the quark singlet superfields.  
$i,j,k$ are the generation indices and we have $\lambda_{ijk} = 
-\lambda_{jik}$, $\lambda^{\prime \prime}_{ijk}
=-\lambda^{\prime \prime}_{ikj}$.  The $SU(2)$
and color indices in Eq. (24) are contracted as follows: $L_iQ_jD_k^c =
(\nu_id_j^\alpha-e_iu_j^\alpha)D_{k\alpha}^c$, etc.  The simultaneous presence
of all three terms in Eq. (25) will imply rapid proton decay, which can be
avoided by setting the $\lambda^{\prime \prime} =0$.  In this case,
baryon number remains an unbroken symmetry while 
lepton number is violated.

There are two types of to $\beta\beta_{0\nu}$ decay in this model. One class
dominantly mediated by heavy gluino exchange\cite{moha86} falls into the 
class of type II contributions discussed in the previous section. The 
dominant 
diagram of this class is ahown in Fig. 3. Detailed evaluation of the nuclear 
matrix element for this class of models has recently been carried out by
Hirsch et. al.\cite{hirsch} and they have found that a very stringent bound
on the following R-violating parameter can be given:
\begin{equation}
\lambda^{\prime}_{111}\leq 4\times 10^{-4}\left({{m_{\tilde{q}}}\over{100
GeV}}\right)^2\left({{m_{\tilde{g}}}\over{100 GeV}}\right)^{1/2}
\end{equation}
It has been recently pointed out by Faessler et al\cite{hirsch} that if
one assumes the dominance of pion exchange in these processes, the limits
$\lambda'_{111}$ becomes more stringent by a factor of 2.

\begin{figure}[htb]
\begin{center}
\epsfxsize=7.5cm
\epsfysize=7.5cm
\mbox{\hskip -1.0in}\epsfbox{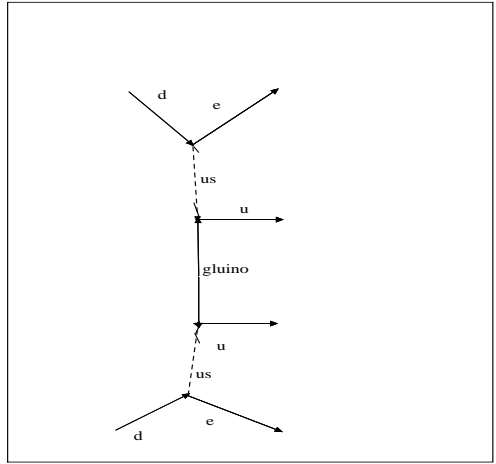}
\caption{  Gluino mediated contribution in MSSM with R-parity violation.
$ds$ stands for the down squark.
\label{Fig.3}} \end{center} 
\end{figure}

The second class of contributions fall into the light neutrino
exchange vector-scalar type\cite{babu} and the dominant diagram of
this type is shown in Fig.4.(where the exchanged scalar particles
 are the $\tilde{b}-\tilde{b}^c$ pair).
This leads to a contribution to $\epsilon_2^{ee}$ given by
\begin{eqnarray}
\epsilon_2^{ee}~\simeq~\left({{(\lambda^{\prime}_{113}}\lambda^{\prime}_{131})
\over{2\sqrt{2}G_F M^2_{\tilde{b}}}}
\right)\left({{ m_b}\over{M^2_{\tilde{b}^c}}}\right)M'
\end{eqnarray}
where $M'=\left(\mu {\rm tan}\beta+A_bm_0\right)$.

\begin{figure}[htb]
\begin{center}
\epsfxsize=6.0cm
\epsfysize=6.0cm
\mbox{\hskip -1.5in}\epsfbox{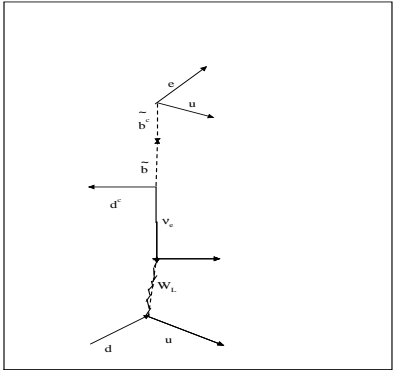}
\caption{ Vector-scalar contribution in MSSM with R-parity violation.
 \label{Fig.4}} \end{center}    
\end{figure}

\noindent Here $A_b,m_0$ are supersymmetry breaking parameters, while $\mu$ is the
supersymmetric mass of the Higgs bosons.  tan$\beta$ is the ratio of the two
Higgs vacuum expectation values and lies in the range $1 \le {\rm tan}\beta
\le m_t/m_b \approx 60$.  
For the choice of all squark masses as well as $\mu$ and the SUSY
breaking mass parameters
being of order of 100 GeV, $A_b=1$, tan$\beta=1$,the following bound
on R-violating couplings is obtained:
\begin{equation}
\lambda^{\prime}_{113}\lambda^{\prime}_{131}\leq 3\times 10^{-8}
\end{equation}
This bound is a more stringent limit on this parameter than
 the existing ones.
The present limits on these parameters are 
$\lambda_{113}' \le 0.03, \lambda_{131}' \le 0.26$, 
which shows that the bound derived here  from $\beta\beta_{0\nu}$ is about
five orders of magnitude more stringent on the product $\lambda_{113}'
\lambda_{131}'$.  If the exchanged scalar particles in Fig.9
are the
$\tilde{s}-\tilde{s}^c$ pair, 
one obtains a limit 
\begin{equation}
\lambda_{121}'\lambda_{112}
\leq 1 \times 10^{-6}
\end{equation}

 which also is  more stringent by about four orders
of magnitude
than the existing limits
($\lambda_{121}' \le 0.26, \lambda_{112}' \le 0.03$).

If the quarks and leptons are composite particles, it is natural to
expect excited leptons which will interact with the electron via some
effective interaction involving the $W_L$ boson. If the excited neutrino
is a majorana particle, then there will be contributions to $\beta\beta_{0\nu}$
decay mediated by the excited neutrinos ($\nu^*$). The effctive interaction
responsible for this is obtained from the primordial interaction:
\begin{equation}
H_{eff}^{\nu^*}= g{{\lambda^{(\nu^*)}_W}\over{ m_{\nu^*}}}\overline{e}
\sigma^{\mu\nu}(\eta^*_L\nu^*_L +\eta^*_R\nu^*_R)W_{\mu\nu}+ h.c.
\end{equation}

Here L and R denote the  left and right chirality states. This contribution
falls into our type B heavy particle exchange category and has been
studied in detail in two recent papers\cite{sriva} and have led to the 
conclusion that it leads to a lower bound 
\begin{equation}
m_{\nu^*}\geq 3.4\times m_W
\end{equation}
for $\lambda^{(\nu^*)}_W\geq 1$. This is a rather stringent bound on 
the compositeness scale.

In conclusion, neutrinoless double beta decay provides a very versatile
way to probe scenarios of physics beyond the standard model. In this review,
we have focussed only on the $0\nu$ mode and briefly touched on the 
single majoron mode. Single and
multi majoron modes which test for the possibility of lepton number being
a spontaneously broken global symmetry have been extensively discussed in 
literature\cite{burgess}.
The $0\nu$ mode acquires special interest in view of the recent 
discoveries in neutrino physics as well as certain SO(10) models
predicting such spectra without contradicting the solar and atmospheric
neutrino data.

\end{document}